\documentclass{article}
\usepackage{amssymb}

\usepackage{amsmath}

\input{tcilatex}

\begin{document}

\title{Continuous-discrete integrable equations and Darboux transformations
as deformations of associative algebras}
\author{B.G.Konopelchenko \\
\\
Dipartimento di Fisica, Universita del Salento \\
and INFN, Sezione di Lecce, 73100, Lecce, Italy}
\maketitle

\begin{abstract}
\bigskip

Deformations of the structure constants for a class of associative
noncommutative algebras generated by Deformation Driving Algebras
(DDA's) are defined and studied. These deformations are governed
by the Central System (CS). Such a CS is studied for the case of
DDA being the algebra of shifts. Concrete examples of deformations
\ for the three-dimensional algebra governed by discrete and mixed
continuous-discrete Boussinesq (BSQ) and WDVV equations are
presented. It is shown that the theory of the Darboux
transformations, at least for the BSQ case, is completely
incorporated into the proposed scheme of deformations.
\end{abstract}

\section{Introduction}

  Deep interrelation between integrable systems and  deformations of the
structure constants for associative algebras  is now well
established due to the discovery of the WDVV equation by Witten
[1] and Dijkgraaf-Verlinde-Verlinde [2], its beautiful
formalization by Dubrovin [3,4] within the theory of Frobenius
manifolds and subsequent extensions (see e.g. [5,6]).

Novel approach to the deformation theory of the structure constants for
associative algebras \ has been proposed recently in [7-11]. \ In contrast
with the theory of Frobenius manifolds and F-manifolds [3-6] where the
algebra action is defined in the tangent bundle the approach [7-11] is
formulated basically in terms of classical and quantum mechanics and
provides us with the coisotropic and quantum deformations of structure
constants. One of its advantages is that it naturally admits different
extensions. To formulate them we recall briefly that of quantum deformations
[10]. It consists basically in three steps: 1) to take the table of
multiplication for an associative algebra in the basis $\mathbf{P}_{0},%
\mathbf{P}_{1},...,\mathbf{P}_{N-1},$ with the structure constants $%
C_{jk}^{l},$ i.e.

\begin{equation}
\mathbf{P}_{j}\mathbf{P}_{k}=C_{jk}^{l}\mathbf{P}_{l},\quad j,k=0,1,...N-1
\end{equation}%
and associate the set of operators

\begin{equation}
f_{jk}=-p_{j}p_{k}+C_{jk}^{l}(x^{0},x^{1},...,x^{N-1})p_{l},\quad
j,k=0,1,...,N-1
\end{equation}%
with it where $x^{0},x^{1},...,x^{N-1}$ stand for the deformation
parameters, $\mathbf{P}_{0}$ denotes the unite element of the basis and
summation over repeated index ( from 0 to N-1 ) is assumed, 2) to require
that the operators $p_{0},p_{1},...,p_{N-1}$and $x^{0},x^{1},...,x^{N-1}$
are elements of the Heisenberg algebra, i.e.

\begin{equation}
\lbrack p_{j},p_{k}]=0, \quad [x^{j},x^{k}]=0, \quad [p_{j},x^{k}]=\hbar
\delta _{j}^{k}, \quad j,k=0,1,...,N-1
\end{equation}
where $\hbar $ is a constant and $\delta _{j}^{k}$ is the Kronecker symbol
and 3) the requirement that the functions $C_{jk}^{l}(x)$ are such that the
set of equations

\begin{equation}
f_{jk}\mid \Psi \rangle =0, \quad j,k=0,1,...,N-1
\end{equation}
has a nontrivial common solution ( Dirac's prescription) where $\mid \Psi
\rangle $ are elements of a certain linear space.

The requirement (4) gives rise to the set of equations ( quantum central
system (QCS)) [10]

\begin{equation}
\hbar \frac{\partial C_{jk}^{n}}{\partial x^{l}}-\hbar \frac{\partial
C_{kl}^{n}}{\partial x^{j}}+C_{jk}^{m}C_{lm}^{n}-C_{kl}^{m}C_{jm}^{n}=0,%
\quad j,k,l,n=0,1,...,N-1
\end{equation}%
which governs quantum deformations of the structure constants $%
C_{jk}^{l}(x^{0},...,x^{N-1})$ . The QCS (5) \ contains the oriented
associativity equation, WDVV equation, Boussinesq equation, Gelfand-Dikii
and Kadomtsev-Petviashvili (KP) hierarchies as the particular cases [10].

\ In this paper we will present an extension of the above aprroach. The
basic idea is to consider other algebras formed by the elements $%
p_{0},...,p_{N-1}$ of the basis and deformation parameters $x^{0},...,x^{N-1}
$ instead of the Heisenberg algebra. We will refer to such algebras as the
Deformation Driving Algebras (DDA's) . We will show that choosing DDA as the
algebra of differences, one gets discrete CS and subsequently discrete
integrable equations and hierarchies. DDA which contains both Heisenberg and
differences subalgebras gives rise to the mixed continuous-discrete
equations and Darboux transformations for continuous systems.

We consider the deformations of the three and four-dimensional \ associative
algebras as the illustration of the general approach. The corresponding CS
contains, in particular, the continuous Boussinesq (BSQ) equation, discrete
and mixed discrete-continuous BSQ equations and corresponding Darboux
transformation. We demonstrate that the Darboux transformations for the BSQ
equation are completely incorporated into the scheme for deformations of
associative algebras presented in the paper.

Another illustrative example is given by the WDVV equation, its discrete and
mixed discrete-continuous versions.

The paper is organized as follows. In section 2 definition of deformations
generated by DDA is given and general CS is derived. Continuous, discrete
and general mixed BSQ deformations are considered in section 3. Section 4 is
devoted to the study of WDVV, discrete and mixed WDVV\ deformations and
equations. Mixed DDA and corresponding Darboux transformations are studied
in section 5. Darboux transformations for the BSQ equation and deformations
for the four-dimensional algebra are discussed in section 6. Appendix
contains some explicit formulas.

\bigskip

\section{Deformations of the structure constants generated by DDA}

\ So, let us consider a finite-dimensional associative algebra \textit{A }%
with (or without) unite element $\mathbf{P}_{0}$ . We choose a basis $%
\mathbf{P}_{0},\mathbf{P}_{1},...,\mathbf{P}_{N-1}$ and introduce structure
constants via the table of multiplication

\begin{equation}
\mathbf{P}_{j}\mathbf{P}_{k}=C_{jk}^{l}\mathbf{P}_{l},\quad j,k=0,1,...,N-1.
\end{equation}

In order to define deformations $C_{jk}^{l}(x^{0},x^{1},...,x^{N-1})$ of the
structure constants we first identify the elements of the basis $P_{j}$ and
deformation parameters $x^{j}$ with the elements of a certain \ algebra
which we will call the Deformation Driving Algebra (DDA) .

Then we introduce the operators

\begin{equation}
f_{jk}\doteqdot -p_{j}p_{k}+C_{jk}^{l}(x)p_{l},\quad j,k=0,1,...,N-1
\end{equation}%
and finally we require that the functions $C_{jk}^{l}(x)$ are such that
these operators have common nontrivial kernel or, equivalently, that the
equations

\begin{equation}
f_{jk}\mid \Psi \rangle =0,\quad j,k=0,1,...,N-1
\end{equation}%
have nontrivial common solutions ( Dirac's prescription).

If the structure constants $C_{jk}^{l}(x)$ are such that this requirement is
satisfied we say that they define deformations generated by DDA.

The requirement (8) can be converted into the set of equations for $%
C_{jk}^{l}(x)$ which we will call the central system (CS) for deformations.
The concrete form of CS depends on DDA. For instance, for the Heisenberg DDA
(3) the CS is given by QCS (5).

\bigskip In the present paper we restrict ourselfs to associative algebras
which have commutative basis, i.e. for which $C_{jk}^{l}(x)=C_{kj}^{l}(x).$
\ For such algebras DDA should have a commutative sublagebra generated by $%
p_{0},...,p_{N-1},$ i.e. $[p_{j},p_{k}]=0$. Examples of such algebras close
to the Heisenberg DDA (3) are given by the well-known algebra of shifts%
\begin{equation}
\lbrack p_{j},p_{k}]=0,\quad \lbrack x^{j},x^{k}]=0,\quad \lbrack
p_{j},x^{k}]=\delta _{j}^{k}p_{j},\quad j,k=0,1,...,N-1
\end{equation}%
algebra of differences

\begin{equation}
\lbrack p_{j},p_{k}]=0,\quad \lbrack x^{j},x^{k}]=0,\quad \lbrack
p_{j},x^{k}]=\delta _{j}^{k}(\hat{I}+p_{j}),\quad j,k=0,1,...,N-1
\end{equation}
algebra of q-differences%
\begin{equation}
\lbrack p_{j},p_{k}]=0,\quad \lbrack x^{j},x^{k}]=0,\quad \lbrack
p_{j},x^{k}]=\delta _{j}^{k}(\hat{I}+qx^{j}p_{j}),\quad j,k=0,1,...,N-1
\end{equation}
and their mixtures. Here $\hat{I}$ denotes the identity operator.

Here we will concentrate on DDA of the types (3),(9), (10) and DDA which
contains them as subalgebras. All these algebras can be presented in the
unified form as the algebra with the following determining commutation
relations

\begin{equation}
\lbrack p_{j},p_{k}]=0,\quad \lbrack x^{j},x^{k}]=0,\quad \lbrack
p_{j},x^{k}]=\delta _{j}^{k}(\hat{I}+\varepsilon _{j}p_{j}),\quad
j,k=0,1,...,N-1
\end{equation}
where $\varepsilon _{j}$ are arbitrary parameters. For $\varepsilon
_{j}\rightarrow 0$ one has the Heisenberg subalgebra while at $\varepsilon
_{j}=1$ one has algebra (10). A realization of such DDA is given by the
algebra of differences $p_{j}=\Delta _{j}.$ where $\Delta _{j}=\frac{1}{%
\varepsilon _{j}}(T_{j}-1)$ and $T_{j}(x^{k})=x^{k}+\delta
_{j}^{k}\varepsilon _{j},j,k=0,1,...,N-1.$ At $\varepsilon _{j}\rightarrow 0$
obviously $\Delta _{j}\rightarrow \frac{\partial }{\partial x^{j}}.$

For the DDA (12) the derivation of the CS is based on the following
identities. The first is

\begin{equation}
\lbrack p_{j},\varphi (x)]=\Delta _{j}\varphi (x)\cdot (\hat{I}+\varepsilon
_{j}p_{j}),\quad j=1,2,...,N-1
\end{equation}%
where $\varphi (x)$ is an arbitrary function, $\Delta _{j}\varphi
(x^{0},x^{1},...,x^{N-1})=(T_{j}-1)\varphi (x^{0},x^{1},...,x^{N-1})$ and $%
T_{j}\varphi (x^{0},...,x^{j},...,x^{N-1})=\varphi
(x^{0},...,x^{j}+\varepsilon _{j},....,x^{N-1}).$ The second is

\begin{eqnarray}
\left( p_{j}p_{k}\right) p_{l}-p_{j}\left( p_{k}p_{l}\right)
=-p_{l}f_{jk}+p_{j}f_{kl}-T_{l}C_{jk}^{m}\cdot f_{lm}+T_{j}C_{kl}^{m}\cdot
f_{jm}+\quad  &&  \notag \\
+(\Delta _{l}C_{jk}^{n}-\Delta _{j}C_{kl}^{n}+T_{l}C_{jk}^{m}\cdot
C_{lm}^{n}-T_{j}C_{kl}^{m}\cdot C_{jm}^{n})p_{n},\quad
j,k,l=1,2,...,N-1.
\end{eqnarray}%
The identity (14) implies that

\begin{eqnarray}
\left\{ \left( p_{j}p_{k}\right) p_{l}-p_{j}\left( p_{k}p_{l}\right)
\right\} \mid \Psi \rangle =A_{klj}^{n}(x)p_{n}\mid \Psi \rangle =\quad  &&
\notag \\
=\left( \Delta _{l}C_{jk}^{n}-\Delta
_{j}C_{kl}^{n}+T_{l}C_{jk}^{m}\cdot
C_{lm}^{n}-T_{j}C_{kl}^{m}\cdot C_{jm}^{n}\right) p_{n}\mid
\Psi\rangle
\end{eqnarray}%
where $\mid \Psi \rangle \subset $ linear subspace $\mathit{H}_{\Gamma }$
defined by equations (8) and $A_{klj}^{n}$ is the associator for the algebra%
\textit{\ A.} \textit{\ }For an associative algebra l.h.s. of (15) vanishes
\ and under the condition that subspace $\mathit{H}_{\Gamma }$ does not
contain elements linear in $p_{j}\mid \Psi \rangle $ equation (15) is
satisfied iff

\begin{equation}
\Delta _{l}C_{jk}^{n}-\Delta _{j}C_{kl}^{n}+T_{l}C_{jk}^{m}\cdot
C_{lm}^{n}-T_{j}C_{kl}^{m}\cdot C_{jm}^{n}=0,\quad j,k,l,n=0,1,...,N-1.
\end{equation}

Thus, we have

\textbf{Proposition. }\ The structure constants $C_{jk}^{l}(x)$ define
deformations generated by the DDA (12) if they obey the CS (16).

We would like to emphasize that the CS (16) is, in fact, the consequence of
the weak associativity condition

\begin{equation}
\left\{ \left( p_{j}p_{k}\right) p_{l}-p_{j}\left( p_{k}p_{l}\right)
\right\} \mid \Psi \rangle =0.
\end{equation}

In standard matrix notations with the matrices $C_{j}$,$A_{lj}$ defined by $%
\left( C_{j}\right) _{k}^{l}=C_{jk}^{l},$ $\left( A_{lj}\right)
_{k}^{n}=A_{klj}^{n}$ the CS (16) takes the form

\begin{equation}
A_{lj}^{d}\doteqdot \Delta _{l}C_{j}-\Delta
_{j}C_{l}+C_{l}T_{l}C_{j}-C_{j}T_{j}C_{l}=0
\end{equation}

or

\begin{equation*}
A_{lj}^{d}\doteqdot (1+\varepsilon _{l}C_{l})T_{l}(1+\varepsilon
_{j}C_{j})-(1+\varepsilon _{j}C_{j})T_{j}(1+\varepsilon _{l}C_{l})=0.
\end{equation*}
There is no summation over repeated indices in these formulae.

In the case of all $\varepsilon _{j}=0,T_{j}=1,\Delta _{j}=\frac{\partial }{%
\partial x^{j}}$ and CS (16) coincides with the QCS ( 5). For all $%
\varepsilon _{j}=1$ one has a pure discrete CS which has been discussed in
[11]. Here we have general mixed CS which will be quite useful in discussion
of relation between continuous and discrete equations and Darboux
transformations as well.

We note also that the results of the paper [11] concerning the relation
between the associator and discrete curvature tensor are valid in the
general case too.

\section{Continuous and discrete Boussinesq equations}

A simple nontrivial example of the proposed scheme corresponds to the
three-dimensional algebra with \ the unite element and the basis $\mathbf{P}%
_{0},\mathbf{P}_{1},\mathbf{P}_{2}.$ The table of multiplication is given by
the trivial part $\mathbf{P}_{0}\mathbf{P}_{j}=\mathbf{P}_{j},j=0,1,2$ and by

\begin{eqnarray}
\mathbf{P}_{1}^{2}
=A\mathbf{P}_{0}+B\mathbf{P}_{1}+C\mathbf{P}_{2},
\notag \\
\mathbf{P}_{1}\mathbf{P}_{2} =D\mathbf{P}_{0}+E\mathbf{P}_{1}+G\mathbf{P}%
_{2}, \\
\mathbf{P}_{2}^{2}
=L\mathbf{P}_{0}+M\mathbf{P}_{1}+N\mathbf{P}_{2}  \notag
\end{eqnarray}%
where the structure constants A,B,...,N depend only on the deformation
parameters $x^{1},x^{2}$ due to the cyclicity of the variable $x^{0}$,
associated with the unite element \ $\mathbf{P}_{0}$. \ It is the easy
consequence of the CS at l=0 and $\ C_{j0}^{l}=\delta _{j}^{l}.$

The CS (16) generated by the DDA (12) in this case is of the form

\begin{eqnarray}
\Delta _{1}D-\Delta _{2}A+AE_{1}+DG_{1}-DB_{2}-LC_{2} =0,  \notag \\
\Delta _{1}E-\Delta _{2}B+D_{1}+BE_{1}+EG_{1}-EB_{2}-MC_{2} =0,  \notag \\
\Delta _{1}G-\Delta _{2}C+CE_{1}+GG_{1}-A_{2}-GB_{2}-NC_{2} =0,  \notag \\
\Delta _{1}L-\Delta _{2}D+AM_{1}+DN_{1}-DE_{2}-LG_{2} =0,  \notag \\
\Delta _{1}M-\Delta _{2}E+L_{1}+BM_{1}+EN_{1}-EE_{2}-MG_{2} =0,  \notag \\
\Delta _{1}N-\Delta _{2}G+CM_{1}+GN_{1}-D_{2}-GE_{2}-NG_{2} =0
\end{eqnarray}%
where $A_{j}\doteqdot T_{j}A$ etc. The CS (20) which governs the
deformations of the structure constants A,B,... is the underdetermined one
and , hence, admits the gauge \ freedom. One of the gauges, namely, $%
B=0,C=1,G=0$ has been discussed in [10] for the quantum deformations and it
was shown that in this case the QCS is reduced to the BSQ equation.

Let us consider the same gauge here. \ The first three equations (20) give

\begin{eqnarray}
L &=&\Delta _{1}D-\Delta _{2}A+AE_{1},  \notag \\
M &=&\Delta _{1}E+D_{1},  \notag \\
N &=&E_{1}-A_{2}
\end{eqnarray}

and the rest of the system (20) takes the form

\begin{eqnarray}
\left( \Delta _{1}^{2}-\Delta _{2}+E_{11}-E_{2}-A_{12}\right) E+2\Delta
_{1}D_{1}-\Delta _{2}A_{1}+A_{1}E_{11} =0,  \notag \\
\left( \Delta _{1}^{2}-\Delta _{2}+E_{11}-E_{2}-A_{12}\right) D+\left(
-\Delta _{1}\Delta _{2}+\Delta _{1}E_{1}+D_{11}\right) A+\Delta _{1}(AE_{1})
=0,  \notag \\
\Delta _{1}(2E_{1}-A_{2})+D_{11}-D_{2} =0.
\end{eqnarray}

The CS (22) defines the Boussinesq \ deformations of the structure constants
in (19) generated by the DDA (12). Indeed, in the case $\varepsilon
_{1}=\varepsilon _{2}=0$, i.e. for the Heisenberg DDA the CS (22) becomes ( $%
A_{x_{j}}=\frac{\partial A}{\partial x^{j}},$ etc)

\begin{eqnarray}
E_{x_{1}x_{1}}-E_{x_{2}}+2D_{x_{1}}-A_{x_{2}} =0,  \notag \\
D_{x_{1}x_{1}}-D_{x_{2}}-A_{x_{1}x_{2}}+AE_{x_{1}}+(AE)_{x_{1}}
=0,  \notag
\\
2E-A =0
\end{eqnarray}%
where all integration constants have been choosen to be equal to zero. Hence
, one has the system

\begin{eqnarray}
\frac{1}{2}A_{x_{1}x_{1}}-\frac{3}{2}A_{x_{2}}+2D_{x_{1}} =0,  \notag \\
D_{x_{1}x_{1}}-D_{x_{2}}-A_{x_{1}x_{2}}+\frac{3}{4}(A^{2})_{x_{1}}
=0.
\end{eqnarray}

Eliminating D, one gets the Boussinesq (BSQ) equation

\begin{equation}
A_{x_{2}x_{2}}+\frac{1}{3}A_{x_{1}x_{1}x_{1}x_{1}}-(A^{2})_{x_{1}x_{1}}=0.
\end{equation}

The BSQ equation (25) defines quantum deformations of the structure
constants in (19) with $B=0,C=1,G=0,L=D_{x_{1}}-A_{x_{2}}+\frac{1}{2}A^{2},M=%
\frac{1}{2}A_{x_{1}}+D,N=-\frac{1}{2}A$ and A, D defined by (24).

In the pure discrete case $\varepsilon _{1}=\varepsilon _{2}=1$ the CS (22)
represents the discrete version of the BSQ system (24). It defines the
discrete deformations of the same structure constants.

There are also the mixed cases. The first is $\varepsilon _{1}=0,\varepsilon
_{2}=1(T_{1}=1,\Delta _{1}=\frac{\partial }{\partial x^{1}},\Delta
_{2}=T_{2}-1)$ . The CS (22) is

\begin{eqnarray}
E_{x_{1}x_{1}}+2D_{x_{1}}-(1+E)\Delta _{2}(E+A) =0,  \notag \\
D_{x_{1}x_{1}}-D_{x_{2}}+AE_{x_{1}}+(AE)_{x_{1}}-\Delta
_{2}A_{x_{1}}-D\Delta _{2}(E+A) =0, \\
(2E-A_{2})_{x_{1}}-\Delta _{2}D_{2} =0.  \notag
\end{eqnarray}

The second case corresponds to $\varepsilon _{1}=1,\varepsilon _{2}=0(\Delta
_{1}=T_{1}-1,T_{2}=1,\Delta _{2}=\frac{\partial }{\partial x^{2}})$ and CS
takes the form

\begin{eqnarray}
\Delta _{1}^{2}E+(E_{11}-E-A_{1})E+A_{1}E_{11}+2\Delta
_{1}D_{1}-(E+A_{1})_{x_{2}} =0,  \notag \\
\Delta _{1}^{2}D+(E_{11}-E-A_{1})D+A\Delta _{1}E_{1}+\Delta
_{1}(AE_{1})+D_{11}A-\Delta _{1}A_{x_{2}}-D_{x_{2}} =0,  \notag \\
\Delta _{1}(2E_{1}-A)+D_{11}-D =0.
\end{eqnarray}

Equations (8) provide us with the linear problems for the CS (22). They are

\begin{eqnarray}
\left( \Delta _{2}-\Delta _{1}^{2}+A\right)\mid\Psi \rangle =0,
\notag
\\
\left( \Delta _{1}\Delta _{2}-E\Delta _{1}-D\right)\mid\Psi
\rangle =0
\end{eqnarray}

or equivalently

\begin{eqnarray}
\left( \Delta _{2}-\Delta _{1}^{2}+A\right)\mid \Psi \rangle =0,
\notag
\\
\left( \Delta _{1}^{3}-(E+A)\Delta _{1}-(D+\Delta
_{1}A)\right)\mid \Psi \rangle =0.
\end{eqnarray}

One can check that the equation $f_{22}\mid \Psi \rangle =0$ is a
consequence of (28).

In the quantum case ($\Delta _{j}=\frac{\partial }{\partial x^{j}})$ the
system (29) is the well-known linear system for the continuous BSQ equation
[12]. We note that in this case the third equation $f_{22}\mid \Psi \rangle
=0$ , i.e. $\left( \Delta _{2}^{2}-L-D\Delta _{1}+\frac{1}{2}A\right) \mid
\Psi \rangle =0$ is the consequence of two equations (29).

In the pure discrete case equations (29) represent the linear problems for
the discrete BSQ equation. The first mixed case considered above can be
treated as the BSQ equation with the discrete time. The second case instead
represents a continuous isospectral flow for the third order difference
problem, i.e. a sort of the difference BSQ equation with continuous time.

\section{WDVV, discrete and continuous-discrete WDVV equations}

Another interesting reduction of the CS (20) corresponds to the constraint \
$C=1,G=0,N=0.$ In this case the CS is

\begin{eqnarray}
\Delta _{1}D-\Delta _{2}A+AE_{1}-DB_{2}-L =0,  \notag \\
\Delta _{1}E-\Delta _{2}B+D_{1}+BE_{1}+EG_{1}-EB_{2}-M =0,  \notag \\
E_{1}-A_{2} =0,  \notag \\
\Delta _{1}L-\Delta _{2}D+AM_{1}-DE_{2} =0,  \notag \\
\Delta _{1}M-\Delta _{2}E+L_{1}+BM_{1}-EE_{2} =0,  \notag \\
M_{1}-D_{2} =0.
\end{eqnarray}

Third and sixth equations (30) imply the existence of the functions U and V
such that
\begin{equation*}
A=U_{1},\quad E=U_{2},\quad D=V_{1},\quad M=V_{2}.
\end{equation*}

Substituting the expressions for L and M given by the first two equations
(30), i.e.

\begin{eqnarray}
L =\Delta _{1}V_{1}-\Delta _{2}U_{1}+U_{1}U_{12}-V_{1}B_{2},  \notag \\
M =\Delta _{1}U_{2}-\Delta _{2}B+V_{11}+BU_{12}-U_{2}B_{2}
\end{eqnarray}

into the rest of the system , one gets the following three equations

\begin{eqnarray}
\Delta _{1}U_{1}-\Delta _{2}B+V_{11}+BU_{12}-U_{2}B_{2}-V_{2} =0, \\
(\Delta _{1}^{2}-\Delta _{2})V_{1}-\Delta _{1}\Delta
_{2}U_{1}+\Delta _{1}(U_{1}U_{12})-\Delta
_{1}(V_{1}B_{2})-V_{1}U_{22}+U_{1}V_{12} =0,
\notag \\
\Delta _{1}(V_{2}+V_{11})-\Delta
_{2}(U_{2}+U_{11})+U_{11}U_{112}-U_{2}U_{22}-V_{11}B_{12}+BV_{12}
=0. \notag
\end{eqnarray}

Solution of this system together with the formulas (31) define deformations
of the structure constants A,B,... generated by the DDA (12).

In the pure continuous case ( $\varepsilon _{1}=\varepsilon _{2}=0$ ) the
system (32) takes the form

\begin{eqnarray}
U_{x_{1}}-B_{x_{2}} =0,  \notag \\
(V_{x_{1}}+U_{x_{2}}+U^{2}-VB)_{x_{1}}-V_{x_{2}} =0, \\
V_{x_{1}}-U_{x_{2}} =0.  \notag
\end{eqnarray}

This system of three conservation laws implies the existence of the function
F such that

\begin{equation}
U=F_{x_{1}x_{1}x_{2}},\quad V=F_{x_{1}x_{2}x_{2}}, \quad
B=F_{x_{1}x_{1}x_{1}}.
\end{equation}

In terms of F the system (33) is

\begin{equation}
F_{x_{2}x_{2}x_{2}}-(F_{x_{1}x_{1}x_{2}})^{2}+F_{x_{1}x_{1}x_{1}}F_{x_{1}x_{2}x_{2}}=0.
\end{equation}

It is the famous WDVV\ equation [1,2]. It is a well-known fact the WDVV
equation describes deformations of the three-dimensional algebra (19) under
the reduction $C=1,G=0,N=0$ [3-10]. So, the system (32) represents the
generalization of the WDVV equation to the case of deformations of the same
algebra generated by DDA (12).

In the pure discrete case $\varepsilon _{1}=\varepsilon _{2}=1$ the system
(32) gives us a pure discrete version of the WDVV equation. In the first
mixed case $\varepsilon _{1}=0,\varepsilon _{2}=1$ we have the system

\begin{eqnarray}
U_{x_{1}}-\Delta _{2}(B+V)-U_{2}\Delta _{2}B =0, \\
(V_{x_{1}}+UU_{2}-VB_{2})_{x_{1}}-\Delta _{2}(V+U_{x_{1}})-VU_{22}+UV_{2}
=0,  \notag \\
(V+V_{2})_{x_{1}}-\Delta _{2}(U+U_{2}+UU_{2})-VB_{2}+BV_{2} =0
\notag
\end{eqnarray}

while in the case $\varepsilon _{1}=1,\varepsilon _{2}=0$ one gets

\begin{eqnarray}
\Delta _{1}U_{1}+B\Delta _{1}U+V_{11}-V-B_{x_{2}} =0, \\
\Delta _{1}(\Delta
_{1}V_{1}-U_{1x_{2}}+U_{1}^{2}-V_{1}B)-V_{1}U+U_{1}V_{1}-V_{1x_{2}}
=0,
\notag \\
\Delta
_{1}(V+V_{11})+U_{11}^{2}-U^{2}-V_{11}B_{1}+BV_{1}-(U+U_{11})_{x_{2}}
=0.  \notag
\end{eqnarray}

Equations $f_{jk}\mid \Psi \rangle =0$ for the system (30) in the coordinate
representation $p_{j}=\Delta _{j}$ have the form

\begin{eqnarray}
\left( \Delta _{2}-\Delta _{1}^{2}+B\Delta _{1}+A\right)\mid\Psi
\rangle
=0,  \notag \\
\left( \Delta _{1}\Delta _{2}-E\Delta _{1}-D\right)\mid\Psi
\rangle =0,
\notag \\
\left( \Delta _{2}^{2}-M\Delta _{1}-L\right)\mid\Psi \rangle =0.
\end{eqnarray}

The system (38) in its turn is equivalent to the following

\begin{eqnarray}
\left( \Delta _{2}-\Delta _{1}^{2}+B\Delta _{1}+A\right)\mid\Psi
\rangle
=0,  \notag \\
\left( \Delta _{1}^{3}-B\Delta _{1}^{2}-(\Delta _{1}B+A_{1}+E)\Delta
_{1}-(\Delta _{1}A+D)\right)\mid\Psi \rangle =0,  \notag \\
\left( \Delta _{2}^{2}-M\Delta _{1}-L\right)\mid\Psi \rangle =0.
\end{eqnarray}

The compatibility condition for the above linear problems are equivalent to
the ''discretized'' WDVV equations (30) or (32). In particular, in the pure
continuous case the problems (38) in the coordinate representation are of
the well-known form [3,4]

\begin{eqnarray}
\Psi _{x_{1}x_{1}}&=&F_{x_{1}x_{1}x_{2}}\Psi
+F_{x_{1}x_{1}x_{1}}\Psi
_{x_{1}}+\Psi _{x_{2}},  \notag \\
\Psi _{x_{1}x_{2}}& =&F_{x_{1}x_{2}x_{2}}\Psi
+F_{x_{1}x_{1}x_{2}}\Psi
_{x_{1}},  \notag \\
\Psi _{x_{2}x_{2}}& =&F_{x_{2}x_{2}x_{2}}\Psi
+F_{x_{1}x_{2}x_{2}}\Psi _{x_{1}}
\end{eqnarray}

or equivalently

\begin{eqnarray*}
\Psi _{x_{2}}& =&\Psi _{x_{1}x_{1}}-F_{x_{1}x_{1}x_{1}}\Psi
_{x_{1}}-F_{x_{1}x_{1}x_{2}}\Psi , \\
\Psi _{x_{1}x_{1}x_{1}}& =&F_{x_{1}x_{1}x_{1}}\Psi
_{x_{1}x_{1}}+(2F_{x_{1}x_{1}x_{2}}+F_{x_{1}x_{1}x_{1}x_{1}})\Psi
_{x_{1}}+(F_{x_{1}x_{2}x_{2}}+F_{x_{1}x_{1}x_{1}x_{2}})\Psi , \\
\Psi _{x_{1}x_{1}x_{1}x_{1}}& =&F_{x_{1}x_{1}x_{1}}\Psi
_{x_{1}x_{1}x_{1}}+(2F_{x_{1}x_{1}x_{2}}+F_{x_{1}x_{1}x_{1}x_{1}})\Psi
_{x_{1}x_{1}}+ \\
&&+(3F_{x_{1}x_{1}x_{1}x_{2}}+F_{x_{1}x_{1}x_{1}x_{1}x_{1}}+F_{x_{1}x_{2}x_{2}})\Psi _{x_{1}}+
\\
&&+(F_{x_{1}x_{1}x_{1}x_{1}x_{2}}+F_{x_{1}x_{1}x_{2}x_{2}}+F_{x_{2}x_{2}x_{2}}-(F_{x_{1}x_{1}x_{2}})^{2}+F_{x_{1}x_{1}x_{1}}F_{x_{1}x_{2}x_{2}})\Psi .
\end{eqnarray*}

The first of these equations represents itself the reduction of the linear
problem

\begin{equation}
\Psi _{x_{2}}=\Psi _{x_{1}x_{1}}+\mathit{V}\Psi _{x_{1}}+\mathit{U}\Psi
\end{equation}%
associated with the generalized KP-modified KP hierarchy. So, the WDVV\
equation is a very special reduction of the generalized KP-mKP hierarchy
stationary with the respect to the times $x_{3},x_{4}$. \ Note that the
second and the third of above equations are equivalent if F obeys WDVV\
equation. For the similar results see [13].

\bigskip

\section{Mixed DDA and Darboux transformations}

\bigskip

For $N\geq 4$ the CS (16) ( or (18)) generated by \ the DDA (12) contains
much more mixed cases between the pure continuous and pure discrete
equations. The situation with N-1 continuous variables and one discrete ,
i.e. when $\varepsilon _{0}=\varepsilon _{1}=...=\varepsilon _{N-2}=0$ and $%
\varepsilon _{N-1}=1$ is of particular interest. The commutation relations
(12) defining the DDA of this type can be presented as

\begin{eqnarray}
\lbrack p_{j},p_{k}] =0,\quad \lbrack x^{j},x^{k}]=0,\quad \lbrack
p_{j},x^{k}]=\hbar \delta _{j}^{k},\quad j,k=0,1,...,N-2, \\
\lbrack p_{j},T] =0,\quad \lbrack x^{j},n]=0,\quad \lbrack
p_{j},n]=0,\quad \lbrack T,n]=T
\end{eqnarray}%
where $T=p_{N-1}+1$ and n is a discrete variable associated with T. The
table of multiplication in the general case is

\begin{eqnarray}
\mathbf{P}_{j}\mathbf{P}_{k}& =&\sum_{l=0}^{N-2}C_{jk}^{l}\mathbf{P}%
_{l}+C_{jk}^{N-1}T,\quad j,k=0,1,...N-2  \notag \\
\mathbf{TP}_{k}& =&\sum_{l=0}^{N-2}C_{N-1k}^{l}\mathbf{P}%
_{l}+C_{N-1k}^{N-1}T,\quad k=0,1,...N-2  \notag \\
\mathbf{T}^{2}& =&\sum_{l=0}^{N-2}C_{N-1N-1}^{l}\mathbf{P}%
_{l}+C_{N-1N-1}^{N-1}T.\quad
\end{eqnarray}%
The change of the basis $p_{N-1}\rightarrow T$ in the table (31) is
performed for the further convenience. Under the constraint $%
C_{jk}^{N-1}=0,j,k=0,1,...,N-2$ such an algebra has \ a special meaning.
Namely, it contains the subalgebra spanned by the elements $%
p_{0},p_{1},...,p_{N-2}$. \ As far as the CS is concerned, one can show that
it separates into two parts, namely,%
\begin{equation}
\frac{\partial C_{j}}{\partial x^{k}}-\frac{\partial C_{k}}{\partial x^{j}}%
+[C_{k},C_{j}]=0,\quad j,k=0,1...,N-2
\end{equation}

and

\begin{equation}
\frac{\partial C_{N-1}}{\partial x^{l}}+C_{l}C_{N-1}-C_{N-1}TC_{l}=0, \quad
l=0,1,...,N-2
\end{equation}%
where $TC_{l}(x^{1},x^{2},...,x^{N-2},n)=C_{l}$ $%
(x^{1},x^{2},...,x^{N-2},n+1).$ Note the difference with the CS (16) in
virtue of the change $p_{N-1}\rightarrow T$. Compatibility conditions for
equations (46) are satisfied due to the fact that the matrices $%
TC_{j}(j=0,1,...,N-2)$ obey equations (45) together with $C_{j}$. \

It is not difficult to show that the system (45) contains a closed subsystem
for the first $(N-2)\times (N-2)$ principal minors $\widetilde{C}_{j}$ of
the matrices $C_{j}$ $((\widetilde{C_{j}}%
)_{k}^{l}=C_{jk}^{l},l,k=0,1,...,N-2)$, i.e.

\begin{equation}
\frac{\partial \widetilde{C}_{j}}{\partial x^{k}}-\frac{\partial \widetilde{C%
}_{k}}{\partial x^{j}}+[\widetilde{C}_{k},\widetilde{C}_{j}]=0,\quad
j,k=0,1...,N-2
\end{equation}%
This subsystem defines an integrable system of differential equations which
coincides with the QCS (5) with $j,k,l,n=0,1,...,N-2$ and the independent
variables $x^{1},x^{2},...,x^{N-2}$. The rest of the system (45) and
equations (46) determine a tranformation $\widetilde{C}_{l}\rightarrow T%
\widetilde{C}_{l}$ which converts solutions of the system (47) into the
solutions of the same system.Thus, equations (45)-(46) define a Darboux
tranformation for the continuous system (47).

We will consider the simplest case N=4 to illustrate this general scheme.
Thus, we consider the four-dimensional associative algebra with the
commutative basis $\mathbf{P}_{0},\mathbf{P}_{1},\mathbf{P}_{2},\mathbf{T}$
. The table of multiplication is given by (nontrivial part)%
\begin{eqnarray}
\mathbf{P}_{1}^{2}
&=&A\mathbf{P}_{0}+B\mathbf{P}_{1}+C\mathbf{P}_{2},
\notag \\
\mathbf{P}_{1}\mathbf{P}_{2}& =&D\mathbf{P}_{0}+E\mathbf{P}_{1}+G\mathbf{P}%
_{2},  \notag \\
\mathbf{P}_{2}^{2}
&=&L\mathbf{P}_{0}+M\mathbf{P}_{1}+N\mathbf{P}_{2},
\notag \\
\mathbf{P}_{1}\mathbf{T}& =&a_{1}\mathbf{P}_{0}+a_{2}\mathbf{P}_{1}+a_{3}%
\mathbf{P}_{2}+a_{4}\mathbf{T,}  \notag \\
\mathbf{P}_{2}\mathbf{T}& =&b_{1}\mathbf{P}_{0}+b_{2}\mathbf{P}_{1}+b_{3}%
\mathbf{P}_{2}+b_{4}\mathbf{T,}  \notag \\
\mathbf{T}^{2}& =&c_{1}\mathbf{P}_{0}+c_{2}\mathbf{P}_{1}+c_{3}\mathbf{P}%
_{2}+c_{4}\mathbf{T.}
\end{eqnarray}%
The matrices $C_{1},C_{2},C_{3}(C_{0}=1)$ are

\begin{equation*}
C_{1}=\left(
\begin{array}{cccc}
0 & A & D & a_{1} \\
1 & B & E & a_{2} \\
0 & C & G & a_{3} \\
0 & 0 & 0 & a_{4}%
\end{array}%
\right) ,C_{2}=\left(
\begin{array}{cccc}
0 & D & L & b_{1} \\
0 & E & M & b_{2} \\
1 & G & N & b_{3} \\
0 & 0 & 0 & b_{4}%
\end{array}%
\right) ,C_{3}=\left(
\begin{array}{cccc}
0 & a_{1} & b_{1} & c_{1} \\
0 & a_{2} & b_{2} & c_{2} \\
0 & a_{3} & b_{3} & c_{3} \\
1 & a_{4} & b_{4} & c_{4}%
\end{array}%
\right) .
\end{equation*}

Equation (47) with $j=1,k=2$ is the pure continuous system (20). The rest of
equations (45) have the following form

\begin{eqnarray}
a_{1x_{2}}-b_{1x_{1}}+Da_{2}+La_{3}-Ab_{2}-Db_{3}+b_{1}a_{4}-a_{1}b_{4}
=0,
\\
a_{2x_{2}}-b_{2x_{1}}+Ea_{2}+Ma_{3}-Bb_{2}-Eb_{3}-b_{1}+b_{2}a_{4}-a_{2}b_{4}
=0,
\\
a_{3x_{2}}-b_{3x_{1}}+Ga_{2}+Na_{3}-Cb_{2}-Gb_{3}+a_{1}+b_{3}a_{4}-a_{3}b_{4}
=0,
\\
a_{4x_{2}}-b_{4x_{1}} =0.
\end{eqnarray}%
Matrix equations (46) are equivalent to 24 scalar differential equations.
They are presented in the Appendix.

Analyzing equations (49)-(52) and those in the Appendix, one concludes that
they are not all independent. For instance, equation (49) coincides with the
difference of equations (87) and (78). Equation (50) coincides with the
difference of equations (88) and (79) while equations (51) and (52) are
similar consequences of equations (89), (80) and equations (90), (81),
respectively. Further , it follows that all these equations, in principle,
allow us to reconstruct $a_{j},b_{j},c_{j}(j=1,2,3,4)$ and finally $TA,TB$
etc in terms of A,B etc.

\section{Darboux transformation for the Boussinesq equation as the discrete
deformation}

\bigskip

Here we will present solution of the above equations for the Boussinesq
reduction $B=0,C=1,G=0$. In this case $L=D_{x_{1}}-A_{x_{2}}+\frac{1}{2}%
A^{2},M=\frac{1}{2}A_{x_{1}}+D,N=-\frac{1}{2}A$ (see section 3 ). Let us
begin with equations (77) and (81) which in the BSQ gauge are of the  form

\begin{eqnarray}
a_{4x_{1}}+a_{4}^{2}-b_{4}-TA =0, \\
b_{4x_{1}}+a_{4}b_{4}-\frac{1}{2}a_{4}TA-TD =0.
\end{eqnarray}%
Equation (52) implies that there exists a function $\varphi $ such that

\begin{equation}
a_{4}=(\ln \varphi )_{x_{1}},\quad b_{4}=(\ln \varphi )_{x_{2}}.
\end{equation}%
Substituting these expressions into (53) and (54), one gets the equations

\begin{eqnarray}
\varphi _{x_{2}} =\varphi _{x_{1}x_{1}}-TA\varphi , \\
\varphi _{x_{1}x_{2}} =\frac{1}{2}TA\varphi _{x_{1}}+TD\varphi .
\end{eqnarray}

These equations are just the linear problems (29) for the continuous BSQ
equation for the shifted solutions TA and TD. \ This observation serves a
lot for the further calculations. To simplify them we consider a special
case for which $a_{3}=1,c_{3}=1.$ \ With such a choice equations (51), (76),
(80), (84), (89) take the form

\begin{eqnarray}
-b_{3x_{1}}+a_{1}+\frac{1}{2}A+b_{3}a_{4}-b_{2}-b_{4} =0, \\
a_{2}+a_{4}-b_{3} =0, \\
b_{3x_{1}}+b_{2}+b_{4}-\frac{1}{2}TA =0, \\
c_{2}+c_{4}-b_{3}-T(a_{2}+a_{4}) =0, \\
a_{1}-\frac{1}{2}A+b_{3}a_{4}-\frac{1}{2}TA =0.
\end{eqnarray}%
\ Rather involved analysis, which we omit, shows that equations (58)-(62) \
and others have a solution

\begin{eqnarray}
a_{1}& =&A-(\ln \widetilde{\Psi })_{x_{1}x_{1}}+(\ln \widetilde{\Psi }%
)_{x_{1}}(\ln \varphi )_{x_{1}},\quad a_{2}=-(\ln \widetilde{(\Psi }\varphi
))_{x_{1}},  \notag \\
a_{3}& =&1,\quad a_{4}=(\ln \varphi )_{x_{1}},  \notag \\
b_{1}& =&D-(\ln \widetilde{\Psi })_{x_{1}x_{2}}+(\ln \widetilde{\Psi }%
)_{x_{1}}(\ln \varphi )_{x_{2}},\quad b_{2}=\frac{1}{2}A-(\ln \varphi
)_{x_{2}},  \notag \\
b_{3}& =&-(\ln \widetilde{\Psi })_{x_{1}},\quad b_{4}=(\ln \varphi
)_{x_{2}},
\\
c_{1}& =&A-(\ln \widetilde{\Psi })_{x_{1}x_{1}}-(\ln \widetilde{\Psi }%
)_{x_{1}}(\ln \varphi )_{x_{1}}+(\ln \widetilde{\Psi })_{x_{1}}\frac{%
T\varphi }{\varphi },  \notag \\
c_{2}& =&-(\ln (\widetilde{\Psi }\varphi ))_{x_{1}}-\frac{T\varphi
}{\varphi },\quad c_{3}=1,\quad c_{4}=\frac{T\varphi }{\varphi }
\notag
\end{eqnarray}%
and

\begin{eqnarray}
TA =A-2(\ln \widetilde{\Psi })_{x_{1}x_{1}}, \\
TD =D+\frac{1}{2}(\ln \widetilde{\Psi
})_{x_{1}x_{1}x_{1}}-\frac{3}{2}(\ln \widetilde{\Psi
})_{x_{1}x_{2}}
\end{eqnarray}%
where $\widetilde{\Psi }$ is a solution of the linear problems for the BSQ
equation

\begin{eqnarray}
\Psi _{x_{2}} =\Psi _{x_{1}x_{1}}-A\Psi ,  \notag \\
\Psi _{x_{1}x_{2}} =\frac{1}{2}A\Psi _{x_{1}}+D\Psi
\end{eqnarray}%
and $\varphi $ solves equations (56-57).

\ The formulae (64) and (65) represent the well-known Darboux transformation
for BSQ equation (23) ( see e.g.[14] ). Recall that in the terms of the $%
\tau -$ function defined via

\begin{equation}
A=-2F_{x_{1}x_{1}},\quad D=-\frac{3}{2}F_{x_{1}x_{2}}+\frac{1}{2}%
F_{x_{1}x_{1}x_{1}}
\end{equation}%
the Darboux transformation \ is a very simple one

\begin{equation}
T\tau =\widetilde{\Psi }\tau .
\end{equation}

Thus , we have constructed deformations of the structure constants from the
table of multiplication (48) parametrized by two continuous variables $%
x^{1},x^{2}$ and one discrete variable n. They are governed by the BSQ
equation (23), formulae (21), (63) and transformations (64),(65). From the
viewpoint of integrable systems the corresponding CS \ represents a
composition of the BSQ equation and its Darboux transformation.

These deformations have an important algebraic property. Using expressions
(63), one can show that the linear problems $f_{jk}\mid \Psi \rangle =0$ are
equivalent to the following one

\begin{eqnarray}
\left( p_{2}-p_{1}^{2}+Ap_{0}\right)\mid\Psi \rangle =0,  \notag \\
\left( p_{1}p_{2}-\frac{1}{2}Ap_{1}-Dp_{0}\right)\mid\Psi \rangle
=0,
\end{eqnarray}%
and

\begin{eqnarray}
(p_{1}-a_{4}p_{0})(T-p_{1}+(\ln \widetilde{\Psi
})_{x_{1}}p_{0})\mid\Psi
\rangle =0,  \notag \\
(p_{2}-b_{4}p_{0})(T-p_{1}+(\ln \widetilde{\Psi
})_{x_{1}}p_{0})\mid\Psi
\rangle =0, \\
(T-c_{4}p_{0})(T-p_{1}+(\ln \widetilde{\Psi
})_{x_{1}}p_{0})\mid\Psi \rangle =0.  \notag
\end{eqnarray}

Factorization of the linear operators in the problems (70) which correspond
to the discrete part of the basic equations $f_{jk}\mid \Psi \rangle =0$ is
the characteristic property of the deformations considered above.

In the coordinate representation for which $p_{1}=\partial
_{x_{1}},p_{2}=\partial _{x_{2}},Tf(x,n)=f(x,n+1)$ the linear problems (70)
have the form

\begin{eqnarray}
(\partial _{x_{1}}-(\ln \varphi )_{x_{1}})\left( T\Psi -(\Psi _{x_{1}}-(\ln
\widetilde{\Psi })_{x_{1}}\Psi )\right) =0,  \notag \\
(\partial _{x_{2}}-(\ln \varphi )_{x_{2}})\left( T\Psi -(\Psi _{x_{1}}-(\ln
\widetilde{\Psi })_{x_{1}}\Psi )\right) =0, \\
(T-\frac{T\varphi }{\varphi })\left( T\Psi -(\Psi _{x_{1}}-(\ln \widetilde{%
\Psi })_{x_{1}}\Psi )\right) =0.  \notag
\end{eqnarray}

These equations imply that

\begin{equation}
T\Psi =\Psi _{x_{1}}-(\ln \widetilde{\Psi })_{x_{1}}\Psi +\alpha \varphi
\end{equation}%
where $\alpha $ is an arbitrary constant. At $\ \alpha =0$ the formula (72)
is nothing but \ the standard transformation of the BSQ wavefunction $\Psi $
under the Darboux transformation \ (see e.g. [14] ). For $\alpha \neq 0$ the
function $T\Psi -\alpha \varphi $ also solves linear problems (66) with the
potentials TA, TD.

Thus, the theory of the Darboux transformations ( at least, for the BSQ
equation) is completely incorporated into the deformation theory presented
in this paper.

The situation with the WDVV\ equation is quite different. At the reduction $%
C=1,G=N=0$ and consequently $E=A,M=D$ one gets equations (49)-(52) and
others which are not too different from those for the BSQ equation. \ But in
contrast with the former one these equations are equivalent to the
overdetermined \ system of nonlinear PDEs which are doubtly compatible. The
conjecture is that in order to construct Darboux type transformation for the
WDVV equation one should consider higher-dimensional algebras ( $N\geq 5).$

\section{Appendix}

\bigskip

In the case N=2 the matrix equation

\begin{equation}
\frac{\partial C_{3}}{\partial x^{1}}+C_{1}C_{3}-C_{3}TC_{1}=0
\end{equation}%
is equivalent to the following twelve equations

\begin{equation}
\frac{\partial a_{1}}{\partial x_{1}}%
+Aa_{2}+Da_{3}+a_{1}a_{4}-a_{1}TB-b_{1}TC=0,
\end{equation}

\begin{equation}
\frac{\partial a_{2}}{\partial x_{1}}%
+a_{1}+Ba_{2}+Ea_{3}+a_{2}a_{4}-a_{2}TB-b_{2}TC=0,
\end{equation}

\begin{equation}
\frac{\partial a_{3}}{\partial x_{1}}%
+Ca_{2}+Ga_{3}+a_{3}a_{4}-a_{3}TB-b_{3}TC=0,
\end{equation}

\begin{equation}
\frac{\partial a_{4}}{\partial x_{1}}+a_{4}^{2}-TA-a_{4}TB-b_{4}TC=0,
\end{equation}

\begin{equation}
\frac{\partial b_{1}}{\partial x_{1}}%
+Ab_{2}+Db_{3}+a_{1}b_{4}-a_{1}TE-b_{1}TG=0,
\end{equation}

\begin{equation}
\frac{\partial b_{2}}{\partial x_{1}}%
+b_{1}+Bb_{2}+Eb_{3}+a_{2}b_{4}-a_{2}TE-b_{2}TG=0,
\end{equation}

\begin{equation}
\frac{\partial b_{3}}{\partial x_{1}}%
+Cb_{2}+Gb_{3}+a_{3}b_{4}-a_{3}TE-b_{3}TG=0,
\end{equation}

\begin{equation}
\frac{\partial b_{4}}{\partial x_{1}}+a_{4}b_{4}-TD-a_{4}TE-b_{4}TG=0,
\end{equation}

\begin{equation}
\frac{\partial c_{1}}{\partial x_{1}}%
+Ac_{2}+Dc_{3}+a_{1}c_{4}-a_{1}Ta_{2}-b_{1}Ta_{3}-c_{1}Ta_{4}=0,
\end{equation}

\begin{equation}
\frac{\partial c_{2}}{\partial x_{1}}%
+c_{1}+Bc_{2}+Ec_{3}+a_{2}c_{4}-a_{2}Ta_{2}-b_{2}Ta_{3}-c_{2}Ta_{4}=0,
\end{equation}

\begin{equation}
\frac{\partial c_{3}}{\partial x_{1}}%
+Cc_{2}+Gc_{3}+a_{3}c_{4}-a_{3}Ta_{2}-b_{3}Ta_{3}-c_{3}Ta_{4}=0,
\end{equation}

\begin{equation}
\frac{\partial c_{4}}{\partial x_{1}}%
+a_{4}c_{4}-Ta_{1}-a_{4}Ta_{2}-b_{4}Ta_{3}-c_{4}Ta_{4}=0.
\end{equation}

Equation

\begin{equation}
\frac{\partial C_{3}}{\partial x^{2}}+C_{2}C_{3}-C_{3}TC_{2}=0
\end{equation}

takes the form

\begin{equation}
\frac{\partial a_{1}}{\partial x_{2}}%
+Da_{2}+La_{3}+b_{1}a_{4}-a_{1}TE-b_{1}TG=0,
\end{equation}

\begin{equation}
\frac{\partial a_{2}}{\partial x_{2}}%
+Ea_{2}+Ma_{3}+b_{2}a_{4}-a_{2}TE-b_{2}TG=0,
\end{equation}

\begin{equation}
\frac{\partial a_{3}}{\partial x_{2}}%
+a_{1}+Ga_{2}+Na_{3}+b_{3}a_{4}-a_{3}TE-b_{3}TG=0,
\end{equation}

\begin{equation}
\frac{\partial a_{4}}{\partial x_{2}}+b_{4}a_{4}-TD-a_{4}TE-b_{4}TG=0,
\end{equation}

\begin{equation}
\frac{\partial b_{1}}{\partial x_{2}}%
+Db_{2}+Lb_{3}+b_{1}b_{4}-a_{1}TM-b_{1}TN=0,
\end{equation}

\begin{equation}
\frac{\partial b_{2}}{\partial x_{2}}%
+Eb_{2}+Mb_{3}+b_{2}b_{4}-a_{2}TM-b_{2}TN=0,
\end{equation}

\begin{equation}
\frac{\partial b_{3}}{\partial x_{2}}%
+b_{1}+Gb_{2}+Nb_{3}+b_{3}b_{4}-a_{3}TM-b_{3}TN=0,
\end{equation}

\begin{equation}
\frac{\partial b_{4}}{\partial x_{2}}+b_{4}^{2}-TL-a_{4}TM-b_{4}TN=0,
\end{equation}

\begin{equation}
\frac{\partial c_{1}}{\partial x_{2}}%
+Dc_{2}+Lc_{3}+b_{1}c_{4}-a_{1}Tb_{2}-b_{1}Tb_{3}-c_{1}Tb_{4}=0,
\end{equation}

\begin{equation}
\frac{\partial c_{2}}{\partial x_{2}}%
+Ec_{2}+Mc_{3}+b_{2}c_{4}-a_{2}Tb_{2}-b_{2}Tb_{3}-c_{2}Tb_{4}=0,
\end{equation}

\begin{equation}
\frac{\partial c_{3}}{\partial x_{2}}%
+c_{1}+Gc_{2}+Nc_{3}+b_{3}c_{4}-a_{3}Tb_{2}-b_{3}Tb_{3}-c_{3}Tb_{4}=0,
\end{equation}

\begin{equation}
\frac{\partial c_{4}}{\partial x_{2}}%
+b_{4}c_{4}-Tb_{1}-a_{4}Tb_{2}-b_{4}Tb_{3}-c_{4}Tb_{4}=0.
\end{equation}

\bigskip

\section{References}

1. Witten E., On the structure of topological phase of two-dimensional
gravity, Nucl. Phys., 340, 281-332 (1990).

2. Dijkgraaf R., Verlinde H. and Verlinde E., Topological strings in d$\prec
$1, Nucl. Phys., B 352, 59-86 (1991).

3. Dubrovin B., Integrable systems in topological field theory, Nucl. Phys.,
B 379, 627-689 (1992).

4. Dubrovin B., Geometry of 2D topological field theories, Lecture Notes in
Math., 1620, 120-348 (1996), Springer, Berlin.

5. Manin Y.I., Frobenius manifolds, quantum cohomology and moduli spaces,
AMS, Providence, 1999.

6. Hertling C., Frobenius manifolds and moduli spaces for singularities,
Cambridge Univ. Press, 2002.

7. Konopelchenko B.G. and Magri F., Coisotropic deformations of associative
algebras and dispersionless integrable hierarchies, Commun. Math. Phys.,
274, 627-658 (2007).

8. Konopelchenko B.G. and Magri F., Dispersionless integrable
equations as coisotropic deformations: extensions and reductions,
Theor. Math. Phys., 151, 803-819 (2007).

9. Konopelchenko B.G. and Magri F., Coisotropic deformations of associative
algebras,Yano manifolds and integrable systems, to appear.

10. Konopelchenko B.G., Quantum deformations of associative algebras and
integrable systems, arXiv:0802.3022,2008.

11. Konopelchenko B.G., Discrete, q-difference deformations of associative
algebras and integrable systems, arXiv:0809.1938, 2008.

12. Zakharov V.E., On stochastization of one-dimensional chain of nonlinear
oscillators, Sov. Phys. JETP, 35, 908-914 (1974).

13. Givental A., $A_{n-1}$ singularities and nKdV hierarchies, Mosc. Math.
J., 3, 475-505 (2003).

14. Matveev V.B. and Salle M.A., Darboux transformations and solitons,
Springer-Verlag, Berlin, 1991.

\bigskip

\bigskip

\end{document}